\documentclass[12pt]{article}
\usepackage{graphics}
\usepackage{amssymb,epsfig,amsmath,euscript,array}
\usepackage{axodraw4j}
\usepackage{subfigure}
\usepackage{color}
\usepackage{cite}
\usepackage{epsfig}
\usepackage{footnote}
\usepackage{caption}

\makeatletter
\@addtoreset{equation}{section}
\makeatother

\newcounter{multieqs}
\newcommand{\ii}{{\rm i}}

\def\bd{\begin{document}}
\def\ed{\end{document}}
\def\nn{\nonumber}
\def\bea{\begin{eqnarray}}
\def\eea{\end{eqnarray}}
\let\bm=\bibitem
\let\la=\label

\begin{document}

\hfill{IPPP/10/92; DCPT/10/184}\\[-0.9cm]

\vspace{20pt}

\begin{center}

{\Large \bf  Resonant Regeneration in the Sub-Quantum Regime} \\[1.5ex]
{\huge \bf --}\\
{\large A demonstration of fractional quantum interference}

\vspace{30pt}

{\bf  John G. Hartnett$^{1}$, Joerg Jaeckel$^{2}$, Rhys G. Povey$^{1}$ and Michael E. Tobar$^{1}$}

$^{1}${\small \em
{School of Physics, University of Western Australia, \\ Crawley W.A., Australia}}\\
$^{2}${\small \em
{Institute for Particle Physics Phenomenology, Durham
University,\\ Durham DH1 3LE, United Kingdom}}

\vspace{10pt}

{\sffamily \tt
}

\vspace{30pt}
\end{center}

\begin{abstract}
Light shining through wall experiments (in the optical as well as in the microwave regime) are a powerful tool to search for light particles coupled very weakly to photons such as axions or extra hidden sector photons. 
Resonant regeneration, where a resonant cavity is employed to enhance the regeneration rate of photons, is one of the most promising
techniques to improve the sensitivity of the next generation of experiments.
However, doubts have been voiced if such methods work at very low regeneration rates where on average the cavity contains less than one photon. In this note we report on a demonstration experiment using a microwave cavity driven with extremely low power, to show that resonant amplification works also in this regime. In accordance with standard quantum mechanics this is a demonstration that interference also works
at the level of less than one quantum.
As an additional benefit this experiment shows that thermal photons inside the cavity cause no adverse effects.
\end{abstract}

\setcounter{page}{0}
\thispagestyle{empty}
\newpage

\section{Introduction}
One of the most intriguing possibility for physics beyond the Standard Model is the existence of new light particles coupled 
only very weakly to the known particles. 
Indeed, such weakly interacting slim particles (WISPs) seem to be a feature of many extensions of the Standard Model
based on field and string 
theory~\cite{Okun:1982xi,Witten:1984dg,Holdom:1985ag,Dienes:1996zr,Abel:2003ue,Conlon:2006tq,Svrcek:2006yi,Abel:2006qt,Abel:2008ai,Arvanitaki:2009fg,Goodsell:2009xc,Goodsell:2010ie}. 
Prominent examples of these are axions and extra U(1) gauge bosons.
Several astrophysical puzzles could be explained by the existence of such 
particles~\cite{Csaki:2001yk,Csaki:2003ef,Mirizzi:2007hr,Hooper:2007bq,DeAngelis:2007dy,Hochmuth:2007hk,Jaeckel:2008fi,Payez:2008pm,Isern:2008nt,DeAngelis:2008sk,Fairbairn:2009zi,Mirizzi:2009nq,Mirizzi:2009aj,Bassan:2010ya} 
giving more than just theoretical motivation for future experimental searches.

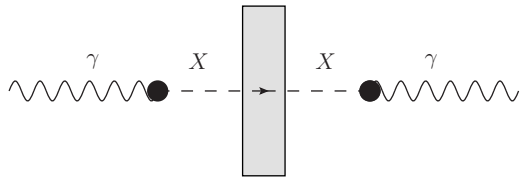
\begin{figure}[t!]
\begin{center}
  \scalebox{0.5}[0.5]{
  \begin{picture}(386,130) (159,-207)
    \SetWidth{1.0}
    \SetColor{Black}
    \GBox(336,-206)(368,-78){0.882}
    \Line[dash,dashsize=8.4,arrow,arrowpos=0.5,arrowlength=5,arrowwidth=2,arrowinset=0.2](272,-142)(432,-142)
    \Photon(160,-142)(272,-142){7.5}{6}
    \Photon(432,-142)(544,-142){7.5}{6}
    \Vertex(432,-142){8}
    \Vertex(432,-142){8}
    \Vertex(272,-142){8}
    \SetOffset(0,-10)
    \Text(224,-110)[c]{\Large{\Black{$\gamma$}}}
    \Text(480,-110)[c]{\Large{\Black{$\gamma$}}}
    \Text(304,-110)[c]{\Large{\Black{$X$}}}
    \Text(400,-110)[c]{\Large{\Black{$X$}}}
  \end{picture}
  }
  \end{center}
\caption{
Schematic of a ``light shining through wall'' experiment.
An incoming photon $\gamma$ is converted into a new particle $X$
which interacts only very weakly with the opaque wall. It passes
through the wall and is subsequently reconverted into an ordinary
photon which can be detected.
}\label{fig:lsw}
\end{figure}

Already the last few years have seen dramatic improvements in laboratory searches. 
Most of this progress~\cite{Ruoso:1992nx,Cameron:1993mr,Robilliard:2007bq,Chou:2007zzc,Pugnat:2007nu,Fouche:2008jk,Afanasev:2008fv,Afanasev:2008jt,Ehret:2009sq,Ehret:2010mh,Povey:2010hs,Wagner:2010mi,Battesti:2010dm} has been obtained using the so-called light-shining-through-walls (LSW)~\cite{Okun:1982xi,Anselm:1986gz,VanBibber:1987rq} technique shown in Fig.~\ref{fig:lsw}.
The idea is as follows. If an
incoming photon is somehow converted into a WISP the latter can
transverse an opaque wall without being stopped. On the other side
of the wall the WISP could then reconvert into a photon\footnote{In principle it does not have to be exactly one WISP traversing the wall. The photon could also convert into a number of virtual WISPs which after the wall recombine into a photon~\cite{Gies:2009wx}. For example this could happen for minicharged particles.}.
The ``light'' doesn't have to be optical but could, for example, be electromagnetic waves at 
microwave frequencies~\cite{Hoogeveen:1992uk,Jaeckel:2007ch,Povey:2010hs,Wagner:2010mi}.

An important next step forward, promising many orders of magnitude improvement in sensitivity, will be the introduction of high quality resonators, i.e. cavities, both in the production and in the regeneration regions~\cite{Hoogeveen:1990vq,Hoogeveen:1992uk,Sikivie:2007qm,Jaeckel:2007ch}. In the production region this has already been pioneered\footnote{It
should be noted that BFRT also used mirrors to enhance the production
probability~\cite{Cameron:1993mr}, but the setup used was an optical delay line and not
a cavity.} by the ALPS experiment~\cite{Ehret:2010mh}.
A high quality factor enhances the production and regeneration probability by the factor of the number of passes, $N_{\rm pass}$, the light makes through the cavity.
On the production side of the experiment this enhancement simply arises due to the fact that the light has a chance to be converted in each pass.
Or, from a different point of view there is simply more light inside the cavity and consequently a higher production rate of WISPs.
On the regeneration side the story is slightly different. After all it is not the driving force, i.e. the WISPs, that is reflected back and forth inside
the cavity. The number of WISPs inside the regeneration side is the same independent of there being a cavity or not.
However, what happens is that the regenerated photons are reflected back and forth inside the cavity and their amplitudes add up
leading to an enhancement proportional to the number of passes.
In turn the total photon power inside the cavity which is proportional to the square of the amplitude inside is enhanced by $N^{2}_{\rm pass}$. 
However, only a fraction of this power $\sim 1/N_{\rm pass}$ leaves the cavity. In total the detectable power output of the cavity is therefore enhanced by $N_{\rm pass}$. 

The enhancement in this ``resonant regeneration'' is based on the positive interference of the photon wave inside the cavity.
Now, one can ask whether this interference still works if the regenerated power is so low that on average there is less than one photon inside the 
cavity?~\cite{questions}
This is the main question addressed in this note.  
We begin with a short theoretical argument and then describe an experiment to demonstrate this sub-quantum interference.
Beyond its purpose to explicitly demonstrate that resonant regeneration works even in the regime where the cavity contains less than one photon
this experimental setup can also be viewed as a nice and simple demonstration of ``less than one photon interfering with itself''.
In a way this experiment is an alternative version of a double slit experiment~\cite{Young:1804} demonstrating interference on the single quantum level.

Finally, it should be noted that as we are only concerned with the regeneration side
the following arguments and demonstration not only apply to LSW experiments but they are, of course, equally applicable to axion dark matter searches~\cite{Sikivie:1983ip,Asztalos:2003px} which rely on an external WISP source and consist of just the regeneration part of an LSW experiment. 

\section{Theoretical considerations}
\subsection{When do we reach the sub-quantum regime?}\label{quantum}
The first question one might have, of course, is if this is really relevant for current or near future LSW experiments.
To see that this is the case let us calculate the power output of a cavity filled with 1 photon.
Let us first consider the ideal case where the only way the cavity loses energy is by photons leaving the cavity towards the detector. 
In this case we have 
\begin{eqnarray}
\label{quantumregime}
P_{\rm out}=P_{\rm loss}=\omega\frac{E_{\rm stored}}{Q}\!\!&=&\!\!\frac{\hbar\omega^2}{Q}
\\\nonumber
\!\!&=&\!\! 4.2\times 10^{-20} {\rm W}\,\left(\frac{f}{\rm GHz}\right)^2\left(\frac{10^5}{Q}\right)
\\\nonumber
\!\!&=&\!\! 1.9\times 10^{-15} {\rm W}\,\left(\frac{\rm \mu m}{\lambda}\right)\left(\frac{10^{5}}{{\mathcal F}}\right)\left(\frac{{\rm m}}{\ell}\right),
\end{eqnarray}
where $Q$ is the quality factor of the cavity\footnote{To be precise the $Q$ here is the loaded quality factor of the cavity. In the idealized
case considered in this section all ``losses'' are due to the photons leaving the cavity towards the detector. Therefore the coupling to the detector determines the loaded $Q$.} and $\omega$ is the angular frequency of the light.
In the second and third line we have calculated the power for typical values for a microwave resonator
of frequency $f$ and for an optical system with wavelength $\lambda$ and a Fabry-Perot cavity of finesse ${\mathcal F}$ and length $\ell$.
In both cases the power at which the sub-quantum regime is reached is well within achievable detector sensitivities.

Therefore the question what happens when the power levels drop into the region when there is less than one quantum
inside the cavity is indeed relevant for future experiments.
Let us now give a brief argument why we think that resonant amplification also works in this regime.
To be specific let us consider an ideal laser generating a plane wave of fixed frequency.
This then turns into a plane wave of WISPs which arrive at our detector cavity where a plane wave of regenerated photons appears.
The plane wave of regenerated photons is a momentum (and energy) eigenstate and as such we have no information on where the photon actually
is. Accordingly interference can happen between different ``parts'' of one photon (and this is all we need for the resonant amplification to occur). 
This is in complete analogy to what happens in a double slit experiment, where one observes interference even when at any given time
only ``half'' of a photon goes through each slit.
 
\subsection{An analogy to the double-slit experiment}
\begin{figure}[t]
\begin{center}
\begin{picture}(200,180)(0,0)
\includegraphics[width=0.4\linewidth]{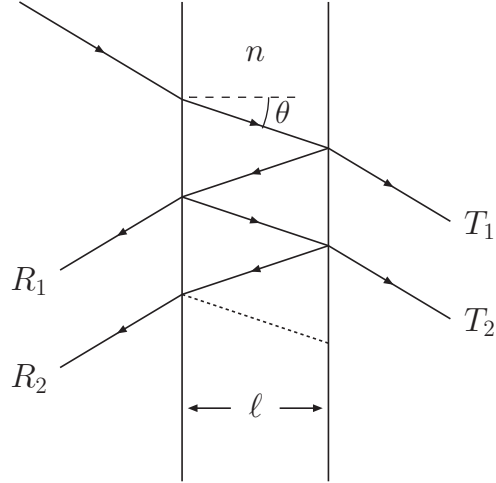}
\DashLine(-115,145)(-75,145){3}
\CArc(-115,145)(30,-23,0)
\Text(-80,140)[c]{$\theta$}
\end{picture}
\end{center}
\caption{\small Light path inside a Fabry-Perot cavity.}
\label{fabry}
\end{figure}

Interference in a cavity is indeed very similar to the famous double-slit experiment~\cite{Young:1804}. To see this let us consider a 
Fabry-Perot cavity as an example (in other resonators the results would be analogous).
Let us shine a plane wave of light at an angle $\theta$ into our cavity as depicted in Fig.~\ref{fabry}.
The total transmitted amplitude will then be the sum of all the transmission amplitudes $T_{k}$ where $k$ denotes the number
of passes through the cavity,
\begin{equation}
\label{geom}
T_{\rm trans}=T_{1}+T_{2}+\ldots=T\sum_{k=0}^{\infty}R^{k} \exp(\ii k \delta)=\frac{T}{1-R\exp(\ii \delta)},
\end{equation}
where $T=(1-R)$ is the transmission amplitude in one pass, $R$ is the amplitude for reflection back to the exit mirror and $\exp(\ii \delta)$ accounts
for the phase accumulated in one pass through the cavity. The transmission probability is given by $|T_{\rm trans}|^2$.

As we can see the transmission amplitude is exactly the sum over all possible paths/passes just as in a double slit experiment. 
The phase difference between the passes is given by
\begin{equation}
\delta=\frac{2\pi}{\lambda}2 n \ell\cos(\theta),
\end{equation}
where $\lambda$ is the incident (vacuum) wavelength, $l$ is the spacing between the mirrors and $n$ is the refractive index.
For real $R$ (pure reflection) close to 1 the transmission is strongly peaked for $\delta=0$. This can be achieved by varying $\theta$ and we obtain
a spatial interference pattern just as in the original double slit experiment.
Alternatively we can, however, also observe the same interference pattern by varying the wavelength, i.e. the frequency.

The higher the reflectivity $R$, i.e. the closer it is to $1$ the stronger the interference pattern is peaked. 
Now, since $N_{\rm pass}\approx 1/(1-R)$, $R$ closer to $1$ means 
that more paths/passes effectively contribute to the geometric sum, Eq.~\eqref{geom}.
In other words the (inverse) width of the interference peak is a measure of the number of paths/passes that interfere.
The transmission probability is given by (for simplicity we use $\cos(\theta)=1$ and $n=1$ from now on),
\begin{eqnarray}
\label{transmission}
\left|T_{\rm trans}\right|^2(\omega)&\!\!=\!\!&\frac{1}{1+\frac{2R}{(1-R)^2}(1-\cos(\delta))}
\\\nonumber
&\!\!\approx\!\!&\frac{1}{1+4\left(\frac{\omega-\omega_{\rm res}}{\omega_{\rm res}}\right)^2\left(\frac{\omega\ell/c}{1-R}\right)^2}
=\frac{1}{1+4Q^2\left(\frac{\omega-\omega_{\rm res}}{\omega_{\rm res}}\right)^2},
\quad{\rm for}\rm \quad \delta,(1-R)\ll1,
\end{eqnarray}
and the width $\Delta \omega$ of this resonance curve is
\begin{equation}
\quad\quad\quad\quad\quad\frac{\Delta\omega}{\omega}=\frac{1-R}{\omega\ell/c}=\frac{1}{N_{\rm pass}\omega\ell/c }=\frac{1}{Q}\quad\quad\quad\quad\quad\quad\quad\quad\quad\,\,\,{\rm for}\quad (1-R)\ll1.
\end{equation}
In the last equalities we have used that the quality factor is also directly related to the width of the resonance curve.

\subsection{Thermal photons and noise}
So far we have completely ignored the issue of thermal noise.
The thermal noise spectrum of a cavity is given by,
\begin{equation}
\label{noiseideal}
\frac{dP_{\rm noise, out}}{d\omega}
=\frac{1}{2\pi}\frac{\hbar\omega}{\exp\left(\frac{\hbar\omega}{k_{B}T}\right)-1}
\frac{1}{1+4Q^{2}\left(\frac{\omega-\omega_{\rm res}}{\omega_{\rm res}}\right)^{2}}.
\end{equation}
Integrating over frequencies we obtain the total noise power coming out of the cavity,
\begin{equation}
P_{\rm noise,out}=\frac{\hbar\omega^{2}_{\rm res}}{Q}\frac{1}{4}\frac{1}{\exp\left(\frac{\hbar\omega_{\rm res}}{k_{B}T}\right)-1},\quad{\rm for}\quad Q\gg1.
\end{equation}
Note that the last factor is basically the Bose-Einstein occupation number of counting the thermal 
photons inside the cavity whereas the first factor is the same as in Eq.~\eqref{quantumregime}. 
Therefore, if the number of photons inside the cavity is large, the thermal noise becomes relevant, even before we reach the regime
where there is less than 1 ``signal'' photon inside the cavity.

The occupation number strongly depends on the frequency and the temperature. In the optical regime
$\hbar\omega\sim 1$~eV. Then even at room temperature, $k_{B}T\sim 1/40$~eV the thermal noise is highly suppressed due to the exponential factor.
However, in the microwave regime  we have $\hbar\omega\sim(1-100)\mu{\rm eV}\ll 1/40{\rm eV}$. Then the number of thermal photons is much
bigger than 1 and thermal noise becomes important even before we reach the subquantum regime. 
Since we performed our measurement in the microwave regime, let us briefly consider this latter case.

The fact that there are actually many thermal photons inside the cavity naturally raises the question whether this (a)~affects the interference in our cavity and, (b)~more practically, can we filter our signal from the thermal noise?

At least theoretically the thermal photons should not cause any problems with the interference. To a very good approximation
photons do not interact with themselves and therefore the presence of thermal photons should not change the behavior 
of our ``signal'' photons. So (a) should not be a problem. Nevertheless, a positive result of our measurement also demonstrates this aspect.

The practical question (b) is more problematic but there is an easy way to solve this issue. If the ``signal'' has a very narrow frequency bandwidth,
then one can search for a signal only at this precise frequency. To do this we basically perform a Fourier transform of the output and study the output power as a function of the frequency (this can be done using a so-called spectrum analyzer; for the specific case of LSW experiments
this technique has been discussed in~\cite{Caspers:2009cj}).
Then we only have to compare our signal to the total noise in this (narrow) frequency bin,
\begin{equation}
\label{noisebin}
P_{\rm noise}=\frac{dP_{\rm noise, out}}{d\omega}\delta \omega,
\end{equation}
where 
\begin{equation}
\delta \omega={\rm Max}(\delta \omega_{\rm resolution},\delta\omega_{signal}),
\end{equation}
and $\delta\omega_{\rm resolution}$ is the frequency resolution of the measurement and $\delta\omega_{\rm signal}$ is the bandwidth of the signal.

Note that the frequency resolution of the measurement is directly related to the measurement time $t_{\rm measure}$,
\begin{equation}
\delta\omega_{\rm resolution}\gtrsim\frac{2\pi}{t_{\rm measure}}.
\end{equation}

In order to measure the transmission probability, Eq.~\eqref{transmission}, at powers in the \linebreak \mbox{(sub-)quantum} regime, Eq.~\eqref{quantumregime}, we need to make sure that the resolution bandwidth is sufficiently small.
In the limit $k_{B}T\gg\hbar\omega$ we need,
\begin{equation}
\label{resolution}
\delta f_{\rm resolution}=\frac{\delta\omega_{\rm resolution}}{2\pi}\leq\frac{\hbar\omega^{2}_{\rm res}}{k_{B}T Q}\left(\frac{S}{N}\right)^{-1}
=10\,{\rm Hz}\left(\frac{f}{\rm GHz}\right)^{2}\left(\frac{300 K}{T}\right)\left(\frac{10^{5}}{Q}\right)\left(\frac{S}{N}\right)^{-1},
\end{equation}
where $S/N$ denotes the desired signal to noise ratio.

\section{Experimental setup and results}
In the previous section we have seen that the small width of the resonance curve arises because a large number of passes interferes which each other.
Indeed the width of the resonance curve is a direct measure of the number of passes effectively contributing to the interference.

As discussed in the introduction the (positive) interference of a large number of passes is what is needed for resonant regeneration.
Therefore, if we find a narrow resonance curve even at very low input power (such that on average there is less than one photon inside the cavity)
we establish this necessary prerequisite for resonant regeneration.

\subsection{Experimental setup}
From the above discussion we see that in order to demonstrate the necessary interference for resonant regeneration we need 
to measure the resonance curve at very low input powers such that on average there is less than one photon inside the cavity.

\begin{figure}[t]
\begin{center}
\includegraphics[angle=0,width=.9\textwidth]{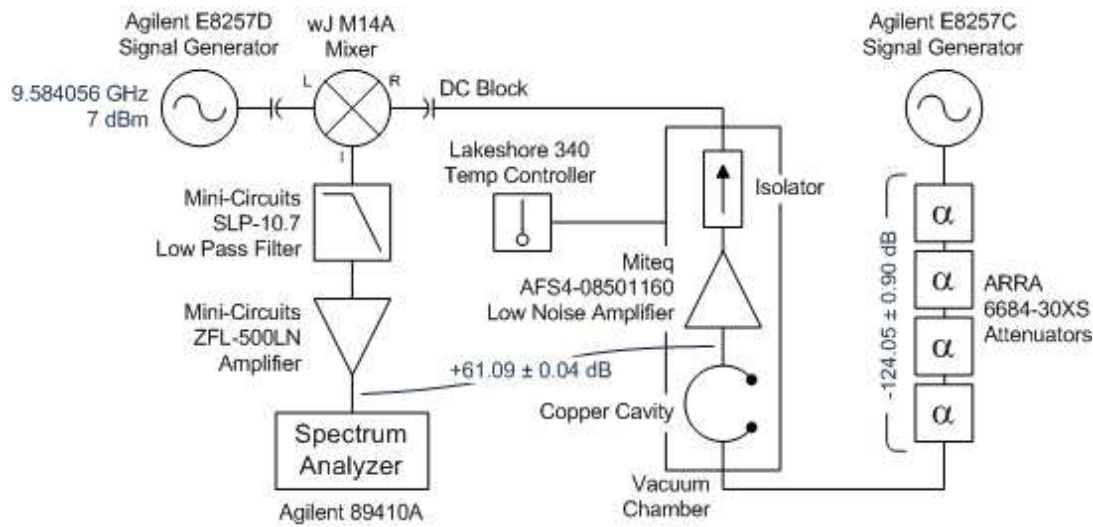}
\end{center}
\vspace{-0cm} \caption{Experimental setup to measure the resonance curve of a cavity at very low input power and correspondingly low power stored inside the cavity. The power output of the generator is fed through a series of attenuators to achieve the desired low incident power into the cavity. The output signal is amplified using a very low noise amplifier and mixed (multiplied) with a fixed frequency signal from a second generator to down convert it to baseband.
The resulting low frequency signal is then filtered, further amplified and analyzed using a FFT spectrum analyzer.}
\label{resonance}
\end{figure}

To do this we used an experimental setup (as shown in Fig.~\ref{resonance}) in the microwave frequency range $f\sim 10\,{\rm GHz}$.
At first glance this seems more difficult
as the power below which the experiment would be in the sub-quantum regime is smaller for lower frequencies (see Sect.~\ref{quantum}).
However, there are two advantages which more than offset this.
Firstly, in the microwave regime even very low power of the order of $10^{-20}\,{\rm W}$ is well within the detectable range.
Secondly, and perhaps even more importantly, microwave generators can be tuned in frequency such that we can easily sweep through 
a band of frequencies around the resonance and measure the resonance curve.

Therefore, we used the setup shown in Fig.~\ref{resonance} which allowed us to measure the resonance curve of a fixed cavity at different levels of input power, corresponding to different numbers of photons inside the cavity.

Our setup (shown in Fig.~\ref{resonance}) consisted of a microwave generator which we used to generate a high resolution signal with variable frequencies between
$9.588$~GHz and $9.593$~GHz. Note, that although the frequency of the generator is variable, once a specific frequency is chosen, the frequency
width of the generated signal is extremely narrow and can be neglected in the analysis.

The generated output goes through a chain of attenuators which reduce the signal power by more than 12 orders of magnitude ($-124$~dB). This allowed us to achieve the very low power levels necessary to probe the quantum regime.
This signal was then fed into a copper cavity (with resonance frequency $f_{\rm res}=9.590$~GHz) for which we intended to observe interference by measuring the resonance curve.

To avoid any drift in the cavity resonance frequency the cavity was placed in a vacuum chamber and kept at a stable temperature $T=305.4\,\mathrm{K}$.
The vacuum chamber also serves as a layer of shielding to avoid picking up external electrical noise.
The transmitted signal is then amplified using a low noise amplifier (which was also located inside the vacuum chamber).
The amplified signal was then mixed with a $9.584$~GHz signal from a second generator and filtered through a low pass filter to obtain a signal in
the MHz range which was analyzed on a FFT spectrum analyzer.

To achieve a sufficient reduction of the thermal noise we need sufficiently good frequency resolution on our spectrum analyzer. For the lowest input
power we chose \linebreak$\delta f_{\rm resolution}=1\,{\rm Hz}$
 
\subsection{Results}
Before analyzing our results we note that to account for the finite coupling of the cavity on the input and the output ports we have to slightly modify Eqs.~\eqref{transmission} and \eqref{quantumregime}. For the transmitted power we have,
\begin{equation}
\label{transmission2}
P_{\rm trans}
=P_{\rm inc}\frac{4\beta_{1}\beta_{2}}{(1+\beta_{1}+\beta_{2})^2}\frac{1}{1+4Q^{\;2}_{L}\left(\frac{\omega-\omega_{\rm res}}{\omega_{\rm res}}\right)^{2}},
\end{equation}
where $\beta_{1}$ and $\beta_{2}$ are the couplings on the input and output ports, respectively, and $Q_L$ is the loaded Q-factor.
The energy stored inside the cavity is given by,
\begin{equation}
\label{stored}
E_{\rm stored}=P_{\rm inc}\frac{Q_{L}}{\omega}\frac{4\beta_{1}}{(1+\beta_{1}+\beta_{2})^2}\frac{1}{1+4Q^{\;2}_{L}\left(\frac{\omega-\omega_{\rm res}}{\omega_{\rm res}}\right)^{2}}.
\end{equation}
\newpage
\begin{figure}[!t]
\begin{center}
\hspace*{0.5mm}\includegraphics[angle=0,width=.703\textwidth]{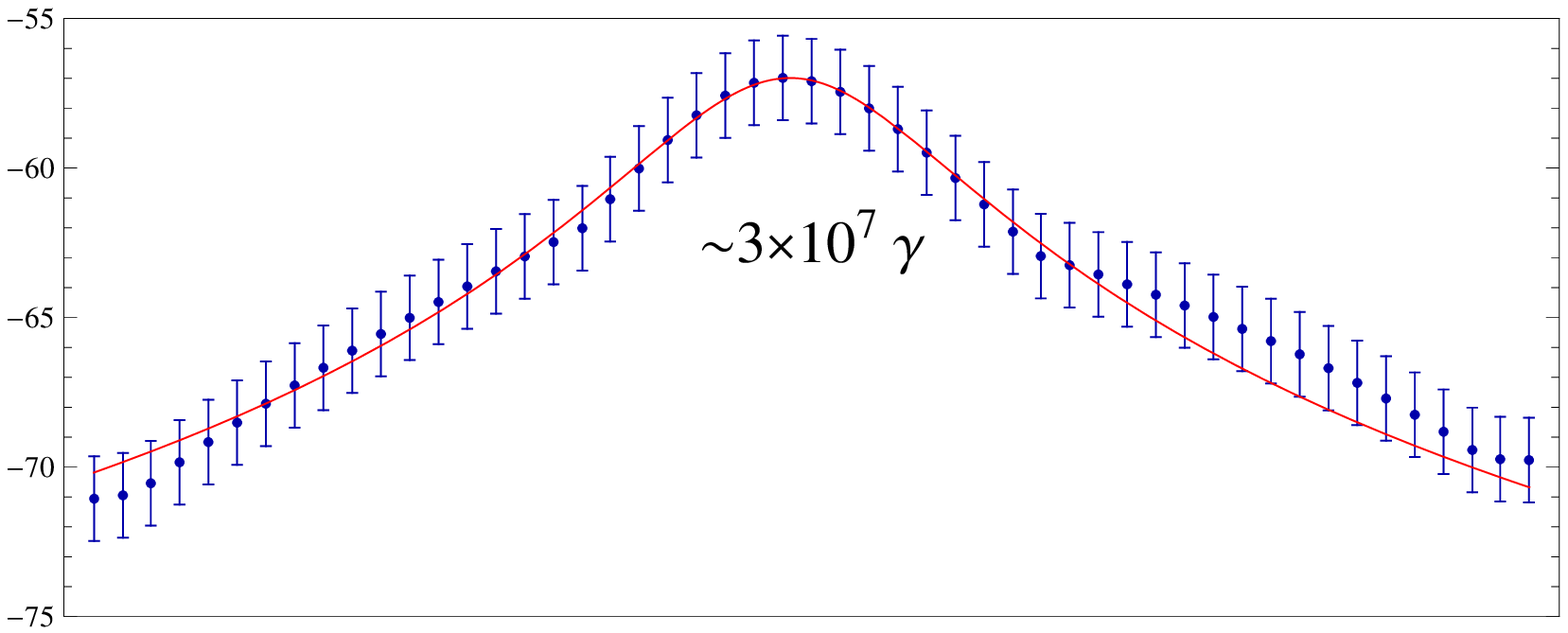}\\
\includegraphics[angle=0,width=.705\textwidth]{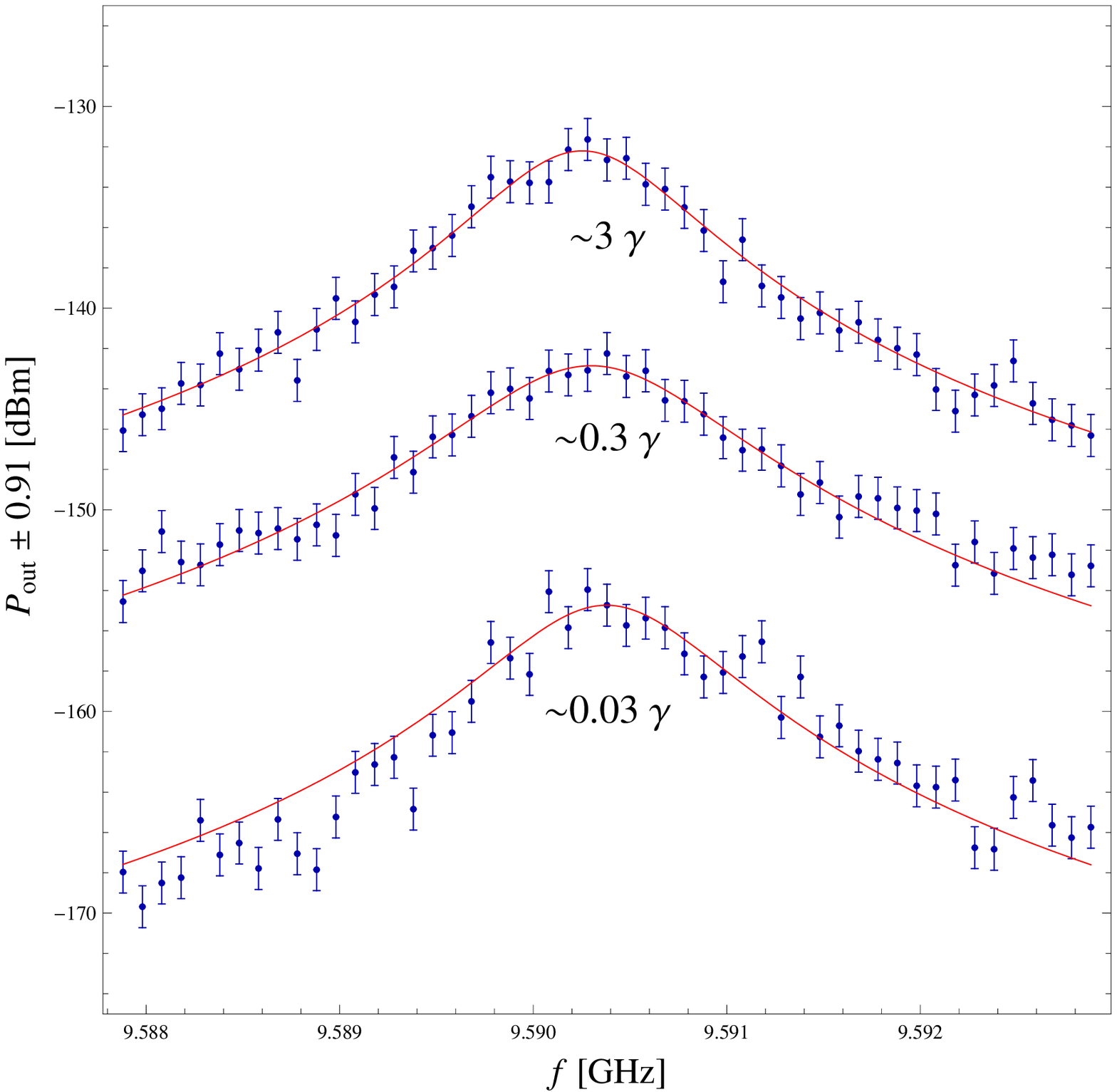}
\end{center}
\vspace{-0cm} \caption{Measured resonance line shapes for various input powers\protect\footnote{}. From top to bottom the input power was $-55$~dBm, $-125$~dBm, $-135$~dBm, and $-145$~dBm.
In the given setup this corresponds to an average of $\sim3\times10^7$, $\sim3$, $\sim0.3$, and $\sim0.03$ photons inside the cavity when the cavity is on resonance. The upper plot gives a check of the classical limit when the input power is high and the number of quanta is large. 
Comparing the lower three curves to this, we can see that the resonance curve has the same shape even when the number of photons inside the cavity is low, thereby demonstrating that interference is present also in this situation.}
\label{resonanceresult}
\end{figure} 
\footnotetext{We note that the output power is somewhat lower than expected from Eq.~\eqref{transmission2}. This is probably due to unaccounted for line losses.}

\begin{table}
\begin{center}
\begin{tabular}{|c|l|c|}
\hline
$P_{\rm in}$ [dBm]& $\#\gamma$ in cavity & $Q_{L}$ \\\hline
-55 & $\quad\sim 3\times10^7$ & 8800\\\hline
-125 & $\quad\sim 3$ & 8900\\\hline
-135 & $\quad\sim 0.3$  & 7100\\\hline
-145 & $\quad\sim 0.03$ & 8200\\\hline
\end{tabular}
\end{center}
\begin{center}
\caption{Measured values for the loaded $Q_{L}$ at different input power. The input coupling coefficient of the cavity was $\beta_{1}=0.89\pm0.05$
and for the output was $\beta_{2}=0.94\pm0.05$. We note that the errors in the determination of $Q_{L}$ are relatively large (of order $10\% - 20\%$) and can be estimated to be around $\Delta Q_{L}\sim 1000$.}
\label{resulttab}
\end{center}
\end{table}
The results of our measurements at different power levels are shown in Fig.~\ref{resonanceresult}.
The upper panel/curve shows a measurement at a relatively large input power of $\sim -55$~dBm$=10^{-8.5}$~W.
Using Eq.~\eqref{stored} we can see that this power corresponds to an averge of approximately $3\times10^7$ photons in the cavity,
certainly in the classical regime.
The three lower curves correspond to lower input powers in the quantum regime: $-125$~dBm, $-135$~dBm, and $-145$~dBm
corresponding to $\sim3$, $\sim0.3$, and $\sim0.03$ photons inside the cavity.
Already on inspection the curves at very low input power are quite similar to the ``classical'' curve suggesting that interference works
as expected.
To check this we have fitted the output power to a Lorentzian curve.
The results of our curve fits are given in Tab.~\ref{resulttab}. As we can see the measured $Q_{L}$ are in reasonable agreement within the uncertainty of $\Delta Q_{L}\sim 1000$ for all power levels. An unmodelled reactive component arising from the coupling probes has somewhat disturbed the Lorentzian line shape.

\begin{figure}[!t]
\begin{center}
\includegraphics[angle=0,width=.7\textwidth]{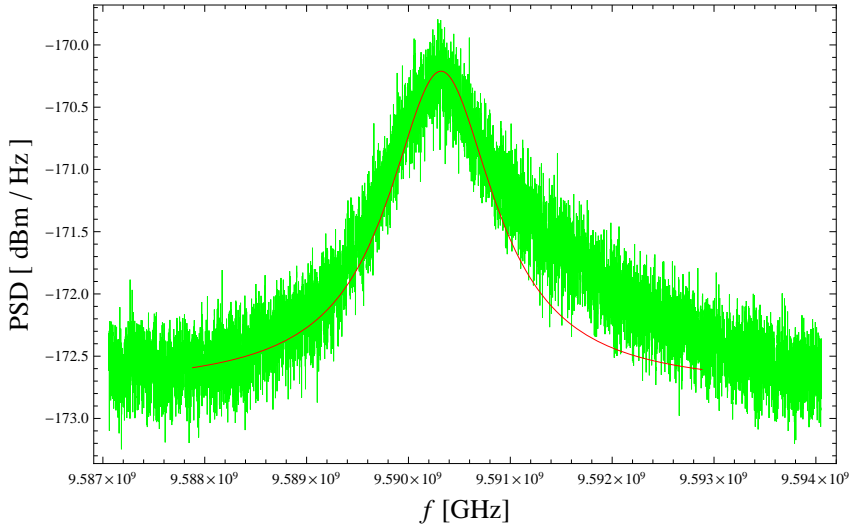}
\end{center}
\vspace{-0cm} \caption{Measurement of the thermal noise spectrum. The frequency resolution was $\delta f_{\rm resolution}=625{\rm Hz}$.
The red fitted curve is a Lorentzian fit with $Q_{L}=6100$.}
\label{noiseresult}
\end{figure}

We have also performed a measurement of the noise spectrum without any input signal.
The result is shown in Fig.~\ref{noiseresult}. We fitted this spectrum with a Lorentzian. The resulting $Q_{L}=6100$ is also in reasonable
agreement with the other measurements.
For this measurement we chose a frequency resolution of $\delta f_{\rm resolution}=625{\rm Hz}$. 
Comparing with Eq.~\eqref{resolution} we see that using this resolution the noise power in each frequency interval actually corresponds to less than one thermal photon inside the cavity per frequency bin. In this sense the thermal noise spectrum itself
can be viewed as a test of the interference with less than one photon inside the cavity.

\section{Conclusions}
In this note we have investigated the question of whether resonant regeneration in light-shining-through-wall experiments still work in
the regime where there is on average less than one photon inside the regeneration cavity.
We have argued that the desired enhancement is directly related to the (positive) interference of the photons in the cavity.
The width of the resonance of a cavity or its quality factor, $Q$, are a direct measure of this interference.
As a demonstration we have measured the $Q$-factor of a microwave cavity at different power levels from the classical
regime, with many photons inside the cavity, to the ``quantum'' regime, with few or even less than one photon inside the cavity.
The measured $Q$ values are in reasonable agreement, demonstrating that the desired degree of interference is present even at very low power levels. This is in agreement with standard quantum mechanical arguments.
Indeed, measuring the transmission curve of a cavity is very similar to a classic double (or better yet, multiple) slit experiment. Interference takes
place between photons doing one pass, two passes and so on inside the cavity. And the transmission curve is the interference pattern in frequency space. In this sense our setup can also be viewed as an alternative version of the double slit experiment.

We have also discussed the role of thermal photons and noise. For frequencies in the microwave range at room temperature the number of thermal photons inside the cavity is larger than 1 since $k_{B}T\gg \hbar\omega$. Our measurements show
that this does not cause any adverse effects to interference, in agreement with the fact that photons only weakly interact with each other.

\section*{Acknowledgements}
J.~J. would like to thank Andrei Afanasev, Holger Gies and Axel Lindner for interesting discussions. Moreover, J.~J. would like to than the 
Frequency Standards and Metrology Research Group at the University of Western Australia for their hospitality.
This work is supported by the Australian Research Council grant DP1092690.

\bibliographystyle{h-physrev5}
\bibliography{../masterbib}
\end{document}